\documentclass{elsarticle}
\usepackage{graphicx}
\usepackage{latexsym}
\usepackage{amsfonts}
\usepackage{amsmath}
\usepackage{amssymb}
\usepackage{revsymb}
\usepackage{bm}
\usepackage{color}

\begin{document}

\newcommand{\be}{\begin{equation}}
\newcommand{\ee}{\end{equation}}
\newcommand{\bea}{\begin{eqnarray}}
\newcommand{\eea}{\end{eqnarray}}
\newcommand{\beq}{\begin{equation}}
\newcommand{\eeq}{\end{equation}}
\newcommand{\beqn}{\begin{eqnarray}}

\newcommand{\eeqn}{\end{eqnarray}}
\newcommand{\ack}[1]{{\bf Pfft! #1}}
\newcommand{\pa}{\partial}
\newcommand{\osigma}{\overline{\sigma}}
\newcommand{\orho}{\overline{\rho}}

\newcommand{\vev}[1]{\langle #1 \rangle}


\title{The Jarzynski Identity and the AdS/CFT Duality}
\author{Djordje Minic\corref{cor1}}
\ead{dminic@vt.edu}
\cortext[cor1]{Corresponding author} 
\author{Michel Pleimling}
\ead{Michel.Pleimling@vt.edu}
\address{Department of Physics, Virginia Tech, Blacksburg, VA 24061, U.S.A.}


\begin{abstract}
We point out a remarkable analogy between the Jarzynski identity from
non-equilibrium statistical physics and the AdS/CFT duality.
We apply the logic that leads to the Jarzynski identity to renormalization group (RG) flows
of quantum field theories and then argue for the natural connection with the
AdS/CFT duality formula. This application can be in principle
checked in Monte Carlo simulations of RG flows.
Given the existing generalizations of the Jarzynski identity in non-equilibrium
statistical physics, and the analogy between the Jarzynski identity and the
AdS/CFT duality, we are led to suggest natural but novel generalizations of
the AdS/CFT dictionary.
\end{abstract}

\maketitle


In this communication we wish to point out a deep analogy between the Jarzynski identity \cite{jarz1,jarz2},
one of the most remarkable results in the recent history of non-equilibrium statistical physics,
and the AdS/CFT duality \cite{adscft1,adscft2,adscft3}, one of the most astonishing developments in
the recent history of quantum field theory and string theory.

The Jarzynski identity has been tested in many experimental situations
in non-equilibrium systems \cite{exp1,exp2,exp3} 
and it has been also theoretically generalized \cite{general1,general2a,general2b,general2c}.
On the other hand, the AdS/CFT duality has been used in fields as diverse as
quantum gravity, quantum chromodynamics, nuclear physics, and condensed
matter physics \cite{adscftapp1,adscftapp2,adscftapp3,adscftapp4,adscftapp5,adscftapp6}. The relationship between non-equilibrium
statistical physics and AdS/CFT duality has been recently discussed 
in the context of aging in systems far from equilibrium \cite{aging1,aging2}.
The present letter aims to establishing a closer connection between these
two fields of physics.

The Jarzynski identity \cite{jarz1,jarz2} gives the {\it exact} relation between 
the thermodynamic free energy differences $\Delta G$ and the irreversible work $W$
\be
\langle \exp{(- \beta W)} \rangle = \exp{(- \beta \Delta G)}
\ee
where $\beta^{-1} = k_B T$ with $k_B$ denoting the Boltzmann constant and 
$T$ the temperature.
The average $\langle...\rangle$ is over all trajectories that take the system from
an initial to the final equilibrium states.
Note that this exact equality extends the well known inequality between
work and change in free energy, $W \ge \Delta G$,  which follows from the second law
of thermodynamics. The relation  $W \ge \Delta G$ is implied by the
Jarzynski identity and Jensen's inequality $\langle e^A\rangle \ge e^{\langle A\rangle}$.

In the AdS/CFT correspondence, one computes the on-shell
bulk action $S_{bulk}$ and relates it to the appropriate boundary
correlators. The {\it conjecture} \cite{adscft1,adscft2,adscft3} is then that
the generating functional of the vacuum correlators of
the operator $O$ for
a $d$ dimensional conformal field theory (CFT) is given by the
partition function $Z(\phi)$ in (Anti-de-Sitter) $AdS_{d+1}$ space
\be
\langle \exp(\int JO ) \rangle = Z (\phi) \to
\exp[- S_{bulk}( {g}, \phi,...)].
\ee
where in the semiclassical limit the
partition function $Z= \exp(-S_{bulk})$.
Here $g$ denotes the metric of the $AdS_{d+1}$ space,
and the boundary values of the bulk field $\phi$ are given by
the sources $J$ of the boundary CFT \cite{note}. Note that we have
written here a semi-classical expression for the correspondence,
which is what is essentially used in many tests of this remarkable
conjecture \cite{adscftapp1,adscftapp2,adscftapp3,adscftapp4,adscftapp5,adscftapp6}.

Obviously, there exists a naive formal similarity between the expressions (1) and (2),
given the fact that $\int JO$ formally corresponds to generalized ``work''.
What we wish to argue in this letter is that this naive similarity is actually deeper
and points to a profound analogy between the two relations.
Given the fact that (1) is exact (under certain assumptions) and (2) is
still regarded as conjectural, but extremely profound and technically powerful,
this analogy might point a way for a formal ``proof'' of (2).
Also, we will argue that this analogy points to some novel views on the
RG flows of quantum field theories as well as to natural generalizations of
the AdS/CFT dictionary.

We begin by stating clearly that we are making two points
in this note:
a) first, after reviewing the proof of Jarzynski's identity in
statistical physics, we propose an analogous identity in the context
of the Wilsonian renormalization group approach, by applying the
same logic applied to the original Jarzynski argument, and by clearly
emphasizing the difference in the respective physics in these two cases, 
and then b) based on the expected
intuitive relation between the Wilsonian renormalization group formalism 
in the boundary quantum field theory and the holographic renormalization in 
the bulk gravitational theory, we propose to use this new form of
the Jarzynksi identity in the context of
AdS/CFT-like duality (with some explicitly stated caveats).
We emphasize that the first point (the RG Jarzynski identity) is very precise, and that the second point concerning the AdS/CFT duality, is essentially heuristic.

First, we start with a path-integral proof of the Jarzynski identity as presented
by Hummer and Szabo \cite{szabo} because we take this formalism to be most
appropriate for the AdS/CFT context.
What Hummer and Szabo pointed out \cite{szabo} is that the Jarzynski
identity follows from the Feynman-Kac theorem for path integrals \cite{stoch1,stoch2}.
In what follows we summarize the argument of Hummer and Szabo and then
we transcribe it to the case of renormalization group (RG) flows of euclidean quantum field theories.

Consider a system whose phase space ($\vec{x}$) density $f(\vec{x}, t)$ evolves according to
the canonical Liouville equation
\be
\frac{\partial f(\vec{x}, t)}{\partial t} = L_t f(\vec{x}, t).
\ee
The Liouville operator $L_t$ explicitly depends on time $t$ and
the Boltzmann distribution is its stationary solution 
\be
L_t e^{-\beta H(\vec{x}, t)} =0.
\ee
Next, Hummer and Szabo consider the unnormalized Boltzmann distribution at time $t$
\be
p(\vec{x}, t) = \frac{e^{-\beta H(\vec{x}, t)} }{\int d\vec{y} e^{-\beta H(\vec{y}, 0)} }.
\ee
This distribution is stationary, i.e. $L_t p(\vec{x}, t) =0$, and it also obviously satisfies
\be
\frac{\partial p}{\partial t} = - \beta (\frac{\partial H}{\partial t}) p.
\ee
Therefore, the above distribution $p(\vec{x}, t)$ is a solution of the following
``sink'' equation, which is of a Fokker-Planck type with a ``sink'' term,
\be
\frac{\partial p}{\partial t}  = L_t p  - \beta (\frac{\partial H}{\partial t}) p.
\ee
Next, Hummer and Szabo point out that the solution of this ``sink'' equation, starting from an equilibrium 
distribution at time $t =0$, can be expressed as a path integral, using the
Feynman-Kac theorem \cite{stoch1,stoch2}, i.e. 
\be
p(\vec{x}, t) = \langle \delta(\vec{x} - \vec{x_t}) \exp [-\beta \int_0^t \frac{\partial H}{\partial t'} (\vec{x_t'}, t') dt']\rangle
\ee
where the average $\langle ... \rangle$ is over an ensemble of trajectories starting from the equilibrium distribution at $t=0$ and evolving according to the Liouville equation.
Each trajectory is weighted with the Boltzmann factor of the external work $W_t$ done
on the system
\be
W_t = \int_0^t \frac{\partial H}{\partial t'} (\vec{x_t'}, t') dt'.
\ee
By remembering that the exponent of the free energy difference is given by definition as
\be
e^{-\beta \Delta G} \equiv \frac{\int d \vec{x} e^{-\beta H(\vec{x}, t) } }{\int d\vec{y} e^{-\beta H(\vec{y}, 0)} }
\ee
we are thus lead to the Jarzynski identity
\be
\exp{(-\beta \Delta G)} \equiv \frac{\int d \vec{x} e^{-\beta H(\vec{x}, t) } }{\int d\vec{y} e^{-\beta H(\vec{y}, 0)} } = \langle \exp{(- \beta W_t)} \rangle .
\ee

In what follows, we will repeat this derivation, step by step, by utilizing the 
well-known formal dictionary between the Hamiltonian $H$ of a dynamical system in phase space
and the action $S$ of a euclidean quantum field theory \cite{wilson1,wilson2}
\be
\beta H(\vec{x}, t) \to S(\varphi, \Lambda)
\ee
where we have also introduced the cut-off $\Lambda$ at which the action of
the quantum field theory is evaluated according to the dynamical RG equation \cite{wilson1,wilson2}.
The RG evolution parameter (``RG time'') is given by the fact that the operation of
rescaling formally corresponds to the ``temporal'' evolution
\be
\Lambda \frac{ \partial}{\partial \Lambda} \to \frac{\partial }{\partial \tau}.
\ee

We wish to be very clear: in the discussion of the Jarzynski-like identity in the context
of the renormalization group we follow the proof just outlined for the case of the
Liouville dynamics, but we point out :
i) that the renormalization group
dynamics is {\it not} the Liouville dynamics, because of its fundamental
irreversible nature and yet ii) the {\it logic} applied to the context
of the Liouville dynamics can be used in the context of the renormalization
group in order to arrive at the statement of the Jarzynski-like
identity! The important point here is that in what follows in the context of
the Wilsonian renormalization group one ultimately gets a stochastic like
equation which is then solved by averaging over the renormalization group
trajectories for the appropriate expressions involving the ``free energy''
and ``work''. This then leads to a new Jarzynski-like identity
involving averages over ensembles of RG (and not dynamical) trajectories!
Here one should emphasize that both the RG ``free energy''
and ``work'' introduced below are defined with respect
to the renormalization group formalism and are fully covariant. Also, this proposed Jarzynski-like
identity can be now tested in numerical RG experiments which
are being planned at the moment.

Thus we can repeat the steps of Hummer and Szabo by replacing $\beta H \to S$ 
and $t \to \tau$.
Therefore, consider a euclidean quantum field theory (and for simplicity, the scalar field theory in 4 space-time dimensions) whose exponent of the effective action $e^{-S(\varphi, \tau)}$ evolves according to
the canonical RG equation \cite{wilson1,wilson2}
\be
\frac{\partial e^{-S_I(\varphi, \tau)}}{\partial \tau} = L_{\tau} e^{-S_I(\varphi, \tau)}
\ee
where the exact RG equation \cite{wilson1,wilson2} states that
\be
L_{\tau} = -\frac{1}{2} \int d^4 p (2 \pi)^4 (p^2 +m^2)^{-1} \frac{\partial K}{\partial \tau} 
\frac{\delta^2}{\delta \varphi(-p) \delta \varphi(p)} .
\ee
Here $K(\frac{p^2}{\Lambda^2})$ is the cut-off function in 4-momentum ($p$) space,
and the total action $S = S_0 +S_I$, where $S_I$ is the interacting part and 
the free part $S_0$ is purely quadratic
\be
S_0 = \frac{1}{2} \int d^4p (2 \pi)^{-4} \varphi(p) \varphi(-p) (p^2 +m^2) K^{-1}(\frac{p^2}{\Lambda^2}).
\ee
The exact RG equation comes from the requirement that 
the generating functional of the vacuum correlators 
$Z[J]_{\tau} = \int D \varphi e^{- S + \int J \varphi} \equiv \langle e^{-\int J \varphi}\rangle_{\tau}$ is $\tau$ independent,
\be
\frac{ \partial Z[J]_{\tau}}{\partial \tau} =0.
\ee
Note that the exact RG equation is equivalent to the fundamental 
Schr\"{o}dinger equation for quantum field theory, or equivalently, to
the knowledge of its kernel, the path integral, as clearly pointed
out in the classic review by Wilson and Kogut 
(pages 154 and 155) \cite{wilson1}.
Thus the RG equation can be precisely viewed as a functional Fokker-Planck
equation for the probability density given by $e^{-S_I}$.
Note that a Fokker-Planck equation with a ``sink'' term is precisely what we have encountered in our 
previous analysis of the Liouville dynamics,
even though the RG and the Liouville operators, and the RG and Liouville 
dynamics are entirely different! The crucial point is that in both cases
we have Fokker-Planck equations, even though generated by
different operators.
Thus, we can draw a precise analogy between an equilibrium distribution and
the conformal fixed point, described by a conformal field theory for which
\be
L_{\tau}  e^{-S_I(\varphi, \tau)} =0.
\ee
Thus in the case of the RG flow, we will consider all trajectories that
connect one conformal fixed point to another one.
This is in complete analogy with all non-equilibrium paths that connect
two equilibrium states, as in the case of the proof of the Jarzynski identity.
Therefore we are led to consider (we use $S$ for $S_I$)
\be
P(\varphi, \tau) = \frac{e^{-S(\varphi, \tau)} }{\int D \psi e^{-S(\psi, \tau_0)} }.
\ee
This distribution is stationary by construction, i.e. $L_{\tau} P =0$, and it also obviously satisfies
\be
\frac{\partial P}{\partial \tau} = - (\frac{\partial S}{\partial \tau}) P.
\ee
Therefore, the above distribution $P$ is a solution of the following
``sink'' equation
\be
\frac{\partial P}{\partial \tau}  = L_{\tau} P - (\frac{\partial S}{\partial \tau}) P.
\ee
For clarity let us stress once again that this is a Fokker-Planck-type equation, as was
the case with the analogous discussion for the Liouville dynamics, even
though the nature of the RG and the Liouville operators that feature in the corresponding
equations is entirely different.
The crucial point is that both equations
can be solved using the same mathematical techniques.
Again, the solution of this ``sink'' equation, starting from an equilibrium 
distribution at time $t =0$ can be expressed as a path integral, using the
Feynman-Kac theorem which applies to any linear stochastic equation (a good reference to
the stochastic equations and the general Feynman-Kac formula is the book by Oeksendal \cite{oeksendal}
as well as the papers by Kac \cite{kac} and Feynman \cite{feynman}):
\be
P(\varphi, \tau) = \langle \delta(\varphi - \varphi_{\tau}) \exp [-\int_0^{\tau} \frac{\partial S}{\partial \tau'} (\varphi_{\tau'}, \tau') d\tau']\rangle
\ee
where the average $\langle ... \rangle$ is over an ensemble of RG trajectories starting from one ``equilibrium distribution``, i.e. one conformal fixed point, which evolves according to the RG equation to another conformal fixed point.
The reason why we average over an ensemble of RG trajectories is that they are
defined by the RG equation and the evolution parameter $\tau$, even though the nature
of the RG trajectories is completely different from the ordinary dynamical trajectories - in
particular the RG flow is not reversible, in contrast to the usual Liouville dynamics. Still, the above
solution is correct because it follows from the general properties of linear stochastic
equations and the Feynman-Kac theorem. Note that in the final result only the ``sink'' term
enters, which was also true in the discussion of the Liouville dynamics.
Each RG trajectory is weighted with the appropriate ``Boltzmann factor'' of the ``external work'' $W_{\tau}$ done
on the system
\be
W_{\tau}= \int_0^\tau \frac{\partial S}{\partial \tau'} (\varphi_{\tau'}, \tau') d\tau'~.
\ee
Also, the ``free energy'' difference is given by definition as
\be
e^{-\Delta G} \equiv \frac{\int D \varphi e^{-S(\varphi, \tau) }}{\int D \psi e^{-S(\psi, \tau_0)} }
\ee
where $\tau_0$ corresponds to the initial cutoff $\Lambda_0$.
We are thus lead to the RG form of the Jarzynski identity
\be \label{RGJ}
\exp{(- \Delta G)} \equiv \frac{\int D \varphi e^{-S(\varphi, \tau)} }{\int D \psi e^{-S(\psi, \tau_0)} } = \langle \exp{(- W_\tau)} \rangle .
\ee
This is the equation that should be tested in numerical RG experiments.
This equations has not been considered in the literature before, even though
it is of an exact form, presumably because the physical construction that leads to
the relevant linear stochastic equation which implies this exact equality,
is not really motivated without thinking about the original Jarzynski equality.

We expect that there exists a natural interpretation of the various quantities involved
in the above discussion of the RG Jarzynski identity, such as the stochastic trajectories between the conformal fixed points and the work done, if one thinks of the specific case
of 2d RG flows. What we have in mind is that the difference between
free energy and work (as defined in terms of the Euclidean
action for the QFT) should define the natural entropy for the 2d flows,
which in turn should be precisely interpreted from the point of view of the c-theorem. In view of our discussion of the RG Jarzynski identity such an interpretation should not be limited to 2d, and
should be applicable to higher dimensional QFTs. We also expect that
the numerical computation should be more efficient for the higher dimensional QFTs as compared to the direct computation. However, due to their special properties 2d QFTs might be exceptional in this regard. 

Next we turn to the relation of the RG formalism and the AdS/CFT
correspondence. 
In this context, the bulk dynamics that is
being rewritten in terms of the so-called holographic renormalization
group formalism is Hamiltonian (i.e. it can be recast in terms of
the Hamiltonian formulation of the bulk gravity), and thus could be understood, in detail, as a
Liouville dynamics! On the other hand, the dual field theory is
understood in the context of the usual Wilsonian RG. The relation
between the two is still on the level of an intuitive, but progressively deeper, understanding \cite{joesb}.
In the previous part of this note we have already applied the logic of the original Jarzynski argument
to the Wilsonian RG and given a proposal for a Jarzynski like identity
in that context. But given the nature of the bulk gravitational
dynamics in the AdS/CFT correspondence
the connection with the Liouville dynamics of the original
Jarzynski argument is much more appropriate. Hence, given our proposal for a Jarzynski-like
identity for the Wilsonian RG flows,  and
the intuitive relation of Wilsonian RG flows and the holographic flows,
as well as the Hamiltonian nature of the latter,
we naturally extrapolate the RG Jarzynski
identity for the vacuum averages which then directly relate
to the fundamental AdS/CFT formula, provided one takes into account the caveats
listed below.
We emphasize that our discussion of the AdS/CFT duality is more speculative
than the application of the Jarzynski identity to the RG. Our arguments regarding 
the AdS/CFT-like duality are
of a heuristic nature, as opposed to the precise statement of the
RG Jarzynski identity.
In particular, to approach the AdS/CFT dictionary from the RG Jarzynski identity we envision an infinitesimal proximity of the initial and final conformal points, with an infinitesimal set of stochastic trajectories connecting them. Then in principle we should have only one CFT to work with, and the averages over the infinitesimal set of stochastic trajectories should be the natural averages pertaining to that CFT, i.e. the averages defined
by the path integral of that CFT. That is why we expect that the
vacuum average should replace the average over RG trajectories.

Next, we equate the work $W_{\tau}$ with the work done by the external source.
This can be understood very simply by invoking the conjugate relation between the
sources and fields with respect to the covariant action. The relation of this
type defines generalized forces (in this cases, sources) and thus 
\be
W_{\tau} \equiv  - \int J \varphi 
\ee
can be understood as a generalized work (where the integral is over space). 
We think that this substitution is natural given the covariant nature
of the RG Jarzynski identity, and the fact that in the case of vacuum
averages, which we have argued replace the average over the RG trajectories, the only ``covariant work'' is done by sourcing the vacuum. We do not see any other natural candidate for such
``covariant work''. 
We also identify the initial and 
the final conformal fixed point and apply the above proposal for the
Jarzynski-like identity in the renormalizaton group context and then we are led to an AdS/CFT-like relation
\be
\langle \exp(\int J \varphi ) \rangle =
\exp (- \Delta G ).
\ee
Note, that we have treated $\varphi$ as the fundamental field.
The same reasoning can be applied to any general operator $O$
in the above euclidean quantum field theory.
Now, we would like to appeal to the extra dimension $\tau$ to argue
that this formula can be rewritten as the actual AdS/CFT relation
provided:

1) We assume a geometrization of assumed conformal invariance in the $\tau$ direction, so
that the metric in the $\tau$ direction has the isometries of the conformal group associated
with the assumed initial and final conformal fixed points.
This leads us to asymptotically AdS metrics
$ds^2 = d r^2 + e^{A r} ds^2_{CFT}$, where $\tau = A r$ in the flat coordinate system 
($A$ determining the size of the bulk space) and
where $ds^2_{CFT}$ is the natural flat metric of the boundary CFT.

2) We assume a map between the choice of RG scheme to the choice of coordinates
in the $\tau$ extra dimensional space, thus effectively inducing gravitational
interactions in this AdS space. This is reasonable from what we know about
perturbative string theory and its relation to the Wilsonian RG \cite{stringrg1,stringrg2,stringrg3}, as well
as from what we know about holographic RG in the context of AdS/CFT \cite{holorg1,holorg2,holorg3,holorg4,holorg5,holorg6,holorg7}.
Nevertheless, this might be harder to justify than the first assumption.

As another general caveat we note that the field theories for which
we expect holographic duals are gauge theories for which we do not have
a nice Wilsonian RG because a cutoff corresponding to a physical length
scale typically breaks gauge invariance, and a cutoff for the gauge theory that could be geometrized is not known at present. 
Also, most field theories do not have a semi-classical
gravity dual and thus AdS/CFT should work only for a limited number
of quantum field theories. Most probably, the theories for which this
duality works can be obtained from the open string sector, in which
case AdS/CFT is really an open/closed string duality of a very specific kind (the gravity dual coming from the closed string sector). A nice discussion of this point is given by
I. Heemskerk and J. Polchinski in the first reference of \cite{joesb}.

Finally, we recall that gravity is a very special interaction whose energy is given in term of boundary data \cite{marolf1,marolf2,marolf3}, or symbolically
\be
\Delta G = \Delta S_{bulk}
\ee
and thus the RG Jarzynski identity, with above assumptions, becomes the canonical AdS/CFT formula.
Note that the semiclassical limit has to come in here, if the expression for the change of the free energy defined in the context of the RG Jarzynski identity is used so that the relative partition function is expanded in some appropriate WKB limit. That WKB expression for the
relative partition function will necessarily involve an exponent of some effective action,
which could be interpreted as an on-shell ``bulk'' action.
Of course, the reason for the fundamental appearance of gravity is obscure in this
heuristic argument. As we have mentioned before, presumably the true origin
of the AdS/CFT duality should be sought in the open/closed string duality, which
would then make the appearance of gravity more palatable.

To conclude: the RG Jarzynski identity is not identical to the usual
Jarzynski identity and the AdS/CFT relation to the RG is illuminated
using the RG Jarzynski identity provided the RG ensemble is replaced
by vacuum averages and provided some natural caveats are met.
Nevertheless, the statement of the RG Jarzynski identity is very precise, and
the connection to the AdS/CFT duality is at this point only heuristic.

This heuristic analogy between the Jarzynski identity and the AdS/CFT duality has many potential applications:
For example, we can envision tests of the RG Jarzynski identity (\ref{RGJ}) that closely mirror the
existing single-molecule tests of the original Jarzynski relation \cite{exp1,exp2,exp3}. In these experiments
a single molecule is repeatedly stretched mechanically and the work is recorded. These non-equilibrium
work fluctuations are then used in order to reconstruct the free-energy landscape of the molecule.
A similar ``pulling'' test of the RG Jarzynski identity can be designed where RG flows are simulated
using Renormalization Group Monte Carlo techniques \cite{mc1,mc2}.
In that case one would measure the generalized work due to the change of the euclidean action
along the RG trajectory and from that reconstruct the partition function for
the euclidean quantum field theory.

Next, we can try to apply the generalizations of Jarzynski's
identity \cite{general1} in order to generalize AdS/CFT-like dualities.
In that case we do not need to assume conformally invariant fixed points.
On the side of non-equilibrium physics \cite{general1},
this would correspond to the situation where one has a probability distribution
of a steady state (ss) with some parameter $\alpha$, $\rho_{ss}(x; \alpha)$, with
the corresponding (negative) ``entropy'' (in the sense of Boltzmann's definition)
\be
\Phi(x; \alpha) = -\log{\rho_{ss}(x; \alpha)}~.
\ee
Given the general properties of probability distributions one
can assert the following mathematical identity \cite{general1}
(for a discrete time evolution, labeled by $i=1,2,...N$)
\be
\langle \prod_{i=0}^{N-1}\frac{\rho_{ss}(x_{i+1}; \alpha_{i+1})}{\rho_{ss}(x_{i+1}; \alpha_{i})}\rangle =1
\ee
that implies in the limit $N \to \infty$ \cite{general1} the generalized Jarzynski identity
\be
\langle \exp(-\int_0^t d t' \frac{d \alpha}{d t'} \frac{\partial \Phi(x;\alpha)}{\partial \alpha})\rangle =1.
\ee
The usual Jarzynski identity follows when $\Phi = -\beta (G-W)$.
Given our dictionary between time and the logarithm of the cut-off  $\Lambda$
($ t \to \tau$) we can obviously translate this general Jarzynski formula into
a general AdS/CFT-like formula, 
\be
\langle \exp(-\int_0^\tau d \tau' \frac{d \alpha}{d \tau'} \frac{\partial \tilde{\Phi}(x;\alpha)}{\partial \alpha})\rangle =1
\ee
which, curiously, involves the gradient of 
``entropy`` $\frac{\partial \tilde{\Phi}}{\partial \alpha}$.
(In the usual AdS/CFT case $\tilde{\Phi} = -(S_{bulk}+\int J O)$.)
This gradient of ``entropy'' corresponds to some kind of ``entropic force'',
a concept that has recently been invoked in the context of the holographic
treatment of gravity \cite{verlinde}. Thus, it is quite plausible that the
concept of entropic force does play a very precise, albeit hidden, role in the AdS/CFT-like
dualities. Such a generalized AdS/CFT formula should be useful in
illuminating the puzzling duals of cosmological backgrounds or pure (non-conformal)
Yang-Mills theory, or various condensed matter systems.

We conclude this letter with the following set of questions:
We have briefly invoked the generalized Jarzynski identity and what this
could imply for the AdS/CFT duality. Conversely, it is natural to ask:
Does the full AdS/CFT duality, in which one uses the 
full bulk partition function instead of $e^{-S_{bulk}}$, say something about even
more generalized versions of
Jarzynski's identity?
In our translation of the Jarzynski identity to the language of RG evolutions of conformal
field theories we have encountered the concept of free energy. On the other hand, the concept
of the c-function is somewhat analogous to free energy. Thus, it is natural to ask: What is the connection of the free energy change $\Delta G$ and
the holographic c-function \cite{holoc}?
Similarly, given the current activity concerning the application of AdS/CFT duality to
many-body physics, one could wonder whether this unexpected relation with the Jarzynski identity illuminates the uses of AdS/CFT in the condensed matter
settings \cite{adscftapp1,adscftapp2,adscftapp3,adscftapp4,adscftapp5,adscftapp6}?
Finally, on a more ambitious and speculative level: Does the topic of this letter point to a more general relation between quantum gravity and
non-equilibrium statistical physics \cite{chia1,chia2,chia3,chia4,chia5,chia6}?
We leave these and many other questions for future work.

\section*{Acknowledgments}
We wish to thank Anne Staples, Attila Szabo and Sinisa Pajevic for inspiring discussions
and P. Arnold, D. Vaman, L. Pando-Zayas and J. Polchinski for comments on the
manuscript. Special thanks to Al Shapere for an illuminating discussion regarding this work and
2d RG flows. We also thank the anonymous referee for very insightful comments on 
the original manuscript.
D. M. acknowledges the hospitality of the National Institutes of Health during the
inception of this project.
D. M.  is supported in part by the U.S.\ Department of Energy under contract DE-FG05-92ER40677.
M. P. acknowledges the support by the US National Science Foundation through grant DMR-0904999.



\begin{thebibliography}{00}

\bibitem{jarz1}
C.~ Jarzynski, Phys. Rev. Lett. {\bf 78}, 2690 (1997). 
\bibitem{jarz2}
C.~ Jarzynski, Phys. Rev. {\bf E 56}, 5018 (1997).

\bibitem{adscft1}
J.~M.~Maldacena,
  Adv.\ Theor.\ Math.\ Phys.\  {\bf 2}, 231 (1998)
  [Int.\ J.\ Theor.\ Phys.\  {\bf 38}, 1113 (1999)].
\bibitem{adscft2}
S.~S.~Gubser, I.~R.~Klebanov and A.~M.~Polyakov,
  Phys.\ Lett.\  B {\bf 428}, 105 (1998).
\bibitem{adscft3}
 E.~Witten,
  Adv.\ Theor.\ Math.\ Phys.\  {\bf 2}, 253 (1998).

\bibitem{exp1}
J.~Liphardt, S.~Dumont, S.~B.~Smith, I.~Tinoco~Jr., and  C.~Bustamante,
  Science {\bf 296}, 1832 (2002).
\bibitem{exp2}
N.~C.~Harris, Y.~Song, and C.-H.~Kiang,
  Phys. Rev. Lett. {\bf 99}, 068101 (2007).
\bibitem{exp3}
A.~Imparato, F.~Sbrana, and M.~Vassalli,
  EPL {\bf 82}, 58006 (2008).

\bibitem{general1}
T.~Hatano and S.-I.~Sasa,
  Phys. Rev. Lett. {\bf 86}, 3463 (2001).
  
\bibitem{general2a}  
R.~J.~Harris and G.~M.~Sch\"{u}tz,
  J. Stat. Mech.: Theory Exp. (2007) P07020.
\bibitem{general2b}
U.~Seifert,
  Eur. Phys. J. B {\bf 64}, 423 (2008).
\bibitem{general2c}
J. Ohkubo,
  J. Phys. Soc. Jpn. {\bf 78}, 123001 (2009).

\bibitem{adscftapp1}
O. Aharony, S.S. Gubser, J. Maldacena, H. Ooguri, Y. Oz,
Phys. Rep. {\bf 323}, 183 (2000).
\bibitem{adscftapp2}
G.~Policastro, D.~T.~Son and A.~O.~Starinets,
  Phys.\ Rev.\ Lett.\  {\bf 87}, 081601 (2001).
\bibitem{adscftapp3}
J.~Erlich, E.~Katz, D.~T.~Son and M.~A.~Stephanov,
  Phys.\ Rev.\ Lett.\  {\bf 95}, 261602 (2005).
\bibitem{adscftapp4}
 S.~A.~Hartnoll, C.~P.~Herzog and G.~T.~Horowitz,
  Phys.\ Rev.\ Lett.\  {\bf 101}, 031601 (2008).
\bibitem{adscftapp5}
D.~T.~Son,
  Phys.\ Rev.\  D {\bf 78}, 046003 (2008).
\bibitem{adscftapp6}
   K.~Balasubramanian and J.~McGreevy,
  Phys.\ Rev.\ Lett.\  {\bf 101}, 061601 (2008).

\bibitem{aging1}
 D.~Minic and M.~Pleimling,
  Phys.\ Rev.\  E {\bf 78}, 061108 (2008).
\bibitem{aging2}
 J.~I.~Jottar, R.~G.~Leigh, D.~Minic and L.~A.~P.~Zayas,
  JHEP {\bf 1011}, 034 (2010).


\bibitem{note}
Essentially, in the language of the second part of this
letter, here one reinterprets the RG flow of the
boundary non-gravitational theory in terms of
bulk gravitational equations of motion, and then
rewrites the generating functional of vacuum correlators
of the boundary theory in terms of a semi-classical
wave function of the bulk ``universe'' with specific boundary
conditions.

\bibitem{szabo}
G. ~Hummer and A.~ Szabo, PNAS {\bf 98}, 3658 (2001).

\bibitem{stoch1}
R.~P.~Feynman, A. R. Hibbs, {\it Quantum Mechanics and Path Integrals},
McGraw-Hill (1965).
\bibitem{stoch2}
Z.~Schuss, {\it Theory and Applications of Stochastic Differential Equations},
Wiley (1980).

\bibitem{wilson1}
K.~G.~Wilson and J.~B.~Kogut,
  Phys.\ Rep.\  {\bf 12}, 75 (1974).
\bibitem{wilson2}
J.~Polchinski,
  Nucl.\ Phys.\  B {\bf 231}, 269 (1984).

\bibitem{oeksendal}
B.~Oeksendal, {\it Stochastic differential equations}, Springer, 5th edition (2000).

\bibitem{kac} M. Kac, Trans. Amer. Math. Soc. {\bf 65}, 1 (1949).

\bibitem{feynman} R. P. Feynman, Rev. Mod. Phys. {\bf 20}, 367 (1948).
  
\bibitem{joesb}
I.~Heemskerk and J.~Polchinski,
arXiv:1010.1264 [hep-th]; I.~Heemskerk, J.~Penedones, J.~Polchinski and J.~Sully,
JHEP {\bf 0910}, 079 (2009);
I.~Heemskerk and J.~Sully, JHEP {\bf 1009}, 099 (2010);
I. Heemskerk, J. Penedones, J. Polchinski, and J. Sully,
JHEP {\bf 0910}, 079 (2009);
A.~L.~Fitzpatrick, E.~Katz, D.~Poland and D.~Simmons-Duffin,
arXiv:1007.2412 [hep-th]; and the references therein.
  
\bibitem{stringrg1}
 T.~Banks and E.~J.~Martinec,
  Nucl.\ Phys.\  B {\bf 294}, 733 (1987).
\bibitem{stringrg2}
   J.~Hughes, J.~Liu and J.~Polchinski,
  Nucl.\ Phys.\  B {\bf 316}, 15 (1989).
\bibitem{stringrg3}
  A.~Polyakov, {\it Gauge Fields and Strings},
  CRC Press (1987).


\bibitem{holorg1}
M. Henningson and K. Skenderis, JHEP {\bf 9807}, 023 (1998).
\bibitem{holorg2}
L.~Susskind and E.~Witten,
  arXiv:hep-th/9805114.
\bibitem{holorg3}
E. Alvarez and C. Gomez,  Nucl. Phys. {\bf B 541}, 441 (1999).
\bibitem{holorg4}
V. Balasubramanian and P. Kraus, Phys. Rev. Lett. {\bf 83},
3605 (1999).
\bibitem{holorg5}
 J.~de Boer, E.~P.~Verlinde and H.~L.~Verlinde,
  JHEP {\bf 0008}, 003 (2000);
  M. Li, Nucl. Phys. B {\bf  579}, 525 (2000).
\bibitem{holorg6}
 V.~Balasubramanian, E.~G.~Gimon and D.~Minic,
  JHEP {\bf 0005}, 014 (2000).
\bibitem{holorg7}
   V.~Balasubramanian, E.~G.~Gimon, D.~Minic and J.~Rahmfeld,
  Phys.\ Rev.\  D {\bf 63}, 104009 (2001).



\bibitem{marolf1}
D.~Marolf,
  Phys.\ Rev.\  D {\bf 79}, 044010 (2009).
\bibitem{marolf2}
D.~Marolf,
  Phys.\ Rev.\  D {\bf 79}, 024029 (2009).
\bibitem{marolf3}
D.~Marolf,
  Gen.\ Rel.\ Grav.\  {\bf 41}, 903 (2009).

\bibitem{mc1}
R.~H.~Swendsen, 
  in {\it Real Space Renormalization}, edited by
  T.~W.~Burkhardt and J.~M.~J.~van~Leeuwen (Springer, Berlin, 1982).
\bibitem{mc2}
R.~Gupta, 
  J. Appl. Phys. {\bf 61}, 3605 (1987).
  
\bibitem{verlinde}
E.~P.~Verlinde,
  arXiv:1001.0785 [hep-th] and references therein.
  For an application of this approach see also,
  C.~M.~Ho, D.~Minic and Y.~J.~Ng,
  Phys. Lett. B {\bf 693}, 567 (2010), and references therein.



\bibitem{holoc}
 D.~Z.~Freedman, S.~S.~Gubser, K.~Pilch and N.~P.~Warner,
  Adv.\ Theor.\ Math.\ Phys.\  {\bf 3}, 363 (1999).


\bibitem{chia1}
D.~Minic and C.~H.~Tze,
  Phys.\ Rev.\  D {\bf 68}, 061501 (2003).
\bibitem{chia2}
D.~Minic and C.~H.~Tze,
  Phys.\ Lett.\  B {\bf 581}, 111 (2004).
\bibitem{chia3}
  V.~Jejjala, D.~Minic,
  Int. J. Mod. Phys. {\bf A22}, 1797 (2007).   
\bibitem{chia4}
  V.~Jejjala, M.~Kavic, D.~ Minic,
  Int. J. Mod. Phys. {\bf A22}, 3317 (2007).
\bibitem{chia5}
  V.~Jejjala, M.~Kavic, D.~Minic and C.~H.~Tze,
  Int. J. Mod. Phys. {\bf A25}, 2515 (2010).
\bibitem{chia6}
  V.~Jejjala, M.~Kavic, D.~Minic and C.~H.~Tze,
  Int. J. Mod. Phys. {\bf D18}, 2257 (2009).

\end{thebibliography}
\end{document}